\pdfoutput=1

\documentclass[11pt]{article}

\usepackage[letterpaper,margin=1in]{geometry}

\usepackage[T1]{fontenc}
\usepackage[utf8]{inputenc}

\usepackage{latexsym,amsfonts,amsmath,amssymb,amstext}

\usepackage{xspace}

\usepackage{lmodern}

\usepackage[final]{microtype}

\usepackage{makeidx}         
\usepackage{graphicx}
\usepackage{caption,subcaption}

\usepackage{multicol}        
\usepackage[bottom]{footmisc}

\usepackage{booktabs,multirow,mathtools}
\usepackage{bm}
\usepackage[version=3]{mhchem}

\usepackage[english]{babel}
\usepackage{csquotes}
\usepackage{textcomp}

\usepackage{siunitx}
\DeclareSIUnit\atm{atm}

\usepackage[retainorgcmds]{IEEEtrantools}

\usepackage{nameref}

\usepackage{verbatim}
\usepackage{listings}
\usepackage{authblk}

\usepackage[hyphens]{url}

\usepackage[linkcolor=blue, citecolor=blue, colorlinks=true, urlcolor=blue, breaklinks=true]{hyperref}
\usepackage[numbers,sort&compress]{natbib}
\usepackage{hypernat}
\usepackage[numbers]{natbib}

\newcommand*{\doi}[1]{\href{https://doi.org/#1}{\nolinkurl{doi:#1}}}

\hypersetup{pdfauthor={Kyle Niemeyer, Chih-Jen Sung},
			pdftitle={GPU-Based Parallel Integration of Large Numbers of Independent ODE Systems}
			}

\begin{document}

\title{GPU-Based Parallel Integration of Large Numbers of Independent ODE Systems\footnote{
	   Final version published in \emph{Numerical Computations with GPUs} (2014) 
	   Springer International Publishing, Ch.~8, pp.~159--182. 
	   V.~Kindratenko (Ed.).~\doi{10.1007/978-3-319-06548-9_8}
	   }}

\author{Kyle E.\ Niemeyer\thanks{Corresponding author; Email: {\tt \href{mailto:kyle.niemeyer@oregonstate.edu}{kyle.niemeyer@oregonstate.edu}}}}
\affil{School of Mechanical, Industrial, \& Manufacturing Engineering, Oregon State University, OR, USA}

\author{Chih-Jen Sung}
\affil{Department of Mechanical Engineering, University of Connecticut, Storrs, CT, USA}

\date{9 June 2014}

\begin{titlepage}
\maketitle

\thispagestyle{empty}

\begin{abstract}
The task of integrating a large number of independent ODE systems arises in various scientific and engineering areas.
For nonstiff systems, common explicit integration algorithms can be used on GPUs, where individual GPU threads concurrently integrate independent ODEs with different initial conditions or parameters.
One example is the fifth-order adaptive Runge--Kutta--Cash--Karp (RKCK) algorithm.
In the case of stiff ODEs, standard explicit algorithms require impractically small time-step sizes for stability reasons, and implicit algorithms are therefore commonly used instead to allow larger time steps and reduce the computational expense.
However, typical high-order implicit algorithms based on backwards differentiation formulae (e.g., VODE, LSODE) involve complex logical flow that causes severe thread divergence when implemented on GPUs, limiting the performance.
Therefore, alternate algorithms are needed.
A GPU-based Runge--Kutta--Chebyshev (RKC) algorithm can handle moderate levels of stiffness and performs significantly faster than not only an equivalent CPU version but also a CPU-based implicit algorithm (VODE) based on results shown in the literature.
In this chapter, we present the mathematical background, implementation details, and source code for the RKCK and RKC algorithms for use integrating large numbers of independent systems of ODEs on GPUs.
In addition, brief performance comparisons are shown for each algorithm, demonstrating the potential benefit of moving to GPU-based ODE integrators.
\end{abstract}

\end{titlepage}

\section{Introduction}
\label{sec:intro}

In a number of science and engineering applications, researchers are faced with the solution of large numbers of independent systems of ordinary differential equations (ODEs).
One prominent example is simulation of reactive flows for modeling combustion~\cite{Knio:1999vd,Day:2000ek,Oran:2001ui,Schwer:2003ws,Ren:2008kd}, atmospheric chemistry~\cite{Alexandrov:1997,Kim:1997}, and groundwater transport~\cite{Barry:1997,Barry:2000}, where operator splitting~\cite{Strang:1968wh,Sportisse:2000gc} decouples the solution of the fluid transport (e.g., advection, diffusion) and chemical kinetics terms.
This results in large numbers of independent ODEs for conservation of chemical species masses, with one system for each spatial location.
The solution of the aggregate of these ODEs consumes most of the total wall-clock time of such simulations, \textgreater90\% in some cases.
Simulations of electrical behavior in cardiac tissue use a similar operator splitting technique, which results in ODE systems for cell membrane dynamics~\cite{Sundnes:2006,Nimmagadda:2011in}.
Other areas that deal with solving many independent systems of ODEs include systems biology~\cite{Dematte:2010kk,Zhou:2011hp} and Monte Carlo methods for sensitivity analysis of ODEs~\cite{Helton:2003,Marino:2008,Kuhn:2009}.

In such problems, large numbers of the same governing ODEs with different initial conditions and\slash or input parameters  must be integrated; since each system is independent, the entire set of ODEs can be integrated concurrently.
While performance can be improved by using parallel central processing unit (CPU) methods, this embarrassingly parallel problem is especially well-suited to acceleration with the thread-based parallelism of graphics processing units (GPUs), as demonstrated for reactive-flow simulations~\cite{Niemeyer:2011uw,Shi:2012cl,Stone:2013gu,Niemeyer:2014dn}.
In particular, Niemeyer and Sung~\cite{Niemeyer:2014dn} recently developed a GPU-based approach for the integration of moderately stiff chemical kinetics ODEs using explicit integration algorithms, using an adaptive fifth-order Runge--Kutta--Cash--Karp (RKCK) method for nonstiff cases and a stabilized second-order Runge--Kutta--Chebyshev (RKC) method for problems with greater stiffness.
For large numbers of ODEs, they demonstrated that the GPU-based RKCK and RKC algorithms performed up to 126 and 65 times faster, respectively, than CPU versions of the same algorithm on a single CPU core.
Furthermore, with moderate levels of stiffness, the GPU-based RKC offered a speedup factor of 57 compared to a conventional implicit algorithm executed in parallel on a six-core CPU.
The specific acceleration demonstrated depending on the problem studied (e.g., the kinetic mechanism) and number of ODEs considered.
Due to the favorable performance of these methods, in this chapter we present the integration algorithms, associated GPU source code, and implementation details so that interested readers can apply them to more general applications (e.g., the areas described above).

\section{Mathematical Background}
\label{sec:math}

In this chapter, we represent a generic system of ODEs using
\begin{equation}
\frac{d \vec{y}}{dt} = \vec{f} \left(t, \vec{y} (t), \vec{g} \right) \;,
\label{eq:ode}
\end{equation}
where $\vec{y} (t)$ is the vector of unknown dependent variables at some time \emph{t}, $\vec{f}$ is the right-hand-side vector function, and $\vec{g}$ is a vector of constant parameters (e.g., pressure or temperature for chemical kinetics).
The size of $\vec{y}$ (i.e., the number of equations\slash unknowns) is \emph{N}.
For the types of problems with which we are concerned here, a large number of ODE systems, $N_{\text{ode}}$, each given by Eq.~\eqref{eq:ode} must be integrated independently from some time $t = t_n$ to $t_{n+1}$, with different initial conditions $\vec{y} \left( t_n \right)$ and constant parameters $\vec{g}$ for each system.
The numerical approximation to the exact solution $\vec{y}(t_n)$ is $\vec{y}_n$, and the step size for a given step is $\delta t_n = t_{n+1} - t_n$.

Nonstiff ODEs, or those with little stiffness, can be solved using explicit integration methods.
Many such methods exist, and algorithms can be classified in general into Runge--Kutta and linear multistep methods, and also into explicit or implicit methods; see Hairer and Wanner~\cite{Hairer:1993} for more details.
Stiffness, a concept somewhat difficult to quantify, refers to the quality of an ODE making standard explicit methods perform poorly due to the requirement for unreasonably small time-step sizes---otherwise the solutions become unstable and oscillate wildly~\cite{Hairer:1996}.

Traditionally, implicit integration algorithms such as those based on backwards difference formulas have been used to handle stiff equations, but these require expensive linear algebra operations on the Jacobian matrix (e.g., LU decomposition, matrix inversion).
The complex logical flow of such operations results in highly divergent instructions across different initial conditions, making implicit algorithms unsuitable for operation on GPUs.
In fact, Stone and Davis~\cite{Stone:2013gu} implemented a traditional high-order implicit algorithm on GPUs, and found that it performed only slightly better than a multi-core CPU version of the same algorithm would.
While implicit algorithms may be required for ODE systems with extreme levels of stiffness (suggesting that new solutions need to be found for GPU acceleration of such problems), other options can be used for cases of little-to-moderate stiffness.
Here, we describe two integration algorithms suitable for use solving many independent systems of ODEs on GPUs.

\subsection{Runge--Kutta--Cash--Karp}
\label{subsec:rkck}

For nonstiff ODEs, an appropriate explicit algorithm is the fifth-order Runge--Kutta method developed by Cash and Karp~\cite{Cash:1990}: the RKCK method.
The RKCK method estimates the local truncation error using an embedded fourth-order method, by taking the difference between the fourth- and fifth-order solutions.
It then uses this estimate to adaptively select the step size~\cite{Press:1992}.

\begin{table}[tbp]
\begin{center}
\begin{tabular}{@{}c c c c c c c c c@{}}
\toprule
\emph{i} & $a_i$ & \multicolumn{5}{c}{$b_{i j}$} & $c_i$ & $c^*_i$ \\ \midrule
1 & &  & & & & & $\frac{37}{378}$ & $\frac{2825}{27648}$ \\
2 & $\frac{1}{5}$ & $\frac{1}{5}$ & & & & & 0 & 0 \\
3 & $\frac{3}{10}$ & $\frac{3}{40}$ & $\frac{9}{40}$ & & & & $\frac{250}{621}$ & $\frac{18575}{48384}$ \\
4 & $\frac{3}{5}$ & $\frac{3}{10}$ & $-\frac{9}{10}$ & $\frac{6}{5}$ & & & $\frac{125}{594}$ & $\frac{13525}{55296}$ \\
5 & 1 & $-\frac{11}{54}$ & $\frac{5}{2}$ & $-\frac{70}{27}$ & $\frac{35}{27}$ & & 0 & $\frac{277}{14336}$ \\
6 & $\frac{7}{8}$ & $\frac{1631}{55296}$ & $\frac{175}{512}$ & $\frac{575}{13824}$ & $\frac{44275}{110592}$ & $\frac{253}{4096}$ & $\frac{512}{1771}$ & $\frac{1}{4}$ \\ \midrule
\multicolumn{2}{c}{\emph{j}} & 1 & 2 & 3 & 4 & 5 & & \\
\bottomrule
\end{tabular}
\caption{Coefficients for the fifth-order Runge--Kutta--Cash--Karp method, adopted from Press et al.~\cite{Press:1992}.}
\label{T:rkck}
\end{center}
\end{table}

Using the terminology established above, the RKCK formulas---which also apply to any general fifth-order Runge--Kutta method---are
\begin{align}
\vec{k_1} &= \delta t_n \, \vec{f} \left( t_n, \vec{y}_n, \vec{g} \right) \;, \label{eq:k1} \\
\vec{k}_2 &= \delta t_n \, \vec{f} \left( t_n + a_2 \, \delta t_n, \vec{y}_n + b_{2 1} \vec{k}_1, \vec{g} \right) \;, \\
\vec{k}_3 &= \delta t_n \, \vec{f} \left( t_n + a_3 \, \delta t_n, \vec{y}_n + b_{3 1} \vec{k}_1 + b_{3 2} \vec{k}_2 , \vec{g}\right) \;, \\
\vec{k}_4 &= \delta t_n \, \vec{f} \left( t_n + a_4 \, \delta t_n, \vec{y}_n + b_{4 1} \vec{k}_1 + b_{4 2} \vec{k}_2 + b_{4 3} \vec{k}_3 , \vec{g} \right) \;, \\
\vec{k}_5 &= \delta t_n \, \vec{f} \left( t_n + a_5 \, \delta t_n, \vec{y}_n + b_{5 1} \vec{k}_1 + b_{5 2} \vec{k}_2 + b_{5 3} \vec{k}_3 + b_{5 4} \vec{k}_4 , \vec{g} \right) \;, \\
\vec{k}_6 &= \delta t_n \, \vec{f} \left( t_n + a_6 \, \delta t_n, \vec{y}_n + b_{6 1} \vec{k}_1 + b_{6 2} \vec{k}_2 + b_{6 3} \vec{k}_3 + b_{6 4} \vec{k}_4 + b_{6 5} \vec{k}_5 , \vec{g} \right) \;, \\
\vec{y}_{n+1} &= \vec{y}_n + c_1 \vec{k}_1 + c_2 \vec{k}_2 + c_3 \vec{k}_3 + c_4 \vec{k}_4 + c_5 \vec{k}_5 + c_6 \vec{k}_6 \;, \\
\vec{y}^*_{n+1} &= \vec{y}_n + c^*_1 \vec{k}_1 + c^*_2 \vec{k}_2 + c^*_3 \vec{k}_3 + c^*_4 \vec{k}_4 + c^*_5 \vec{k}_5 + c^*_6 \vec{k}_6 \label{eq:rkc-ynp1} \;,
\end{align}
where $\vec{y}_{n+1}$ is the fifth-order solution and $\vec{y}^*_{n+1}$ is the solution of the embedded fourth-order method.
The RKCK coefficients are given in Table~\ref{T:rkck}.
The local error $\bm{\varDelta}_{n+1}$ is estimated using the difference between the fourth- and fifth-order solutions:
\begin{equation}
\bm{\varDelta}_{n+1} = \vec{y}_{n+1} - \vec{y}^*_{n+1} = \sum_{i = 1}^6 \left( c_i - c^*_i \right) \vec{k}_i \;.
\end{equation}
Then, this error is compared against a desired accuracy, $\bm{\varDelta_0}$, defined by
\begin{equation}
\bm{\varDelta}_0 = \varepsilon \left( | \vec{y}_n | + \left| \delta t_n \, \vec{f} \left(t_n, \vec{y}_n , \vec{g} \right) \right| + \delta \right) \;,
\end{equation}
where $\varepsilon$ is a tolerance level and $\delta$ represents a small value (e.g., \num{e-30}).
When the estimated error rises above the desired accuracy ($ \bm{\varDelta}_{n+1} > \bm{\varDelta}_0 $), the algorithm rejects the current step and calculates a smaller step size.
Correspondingly, the algorithms accepts a step with error at or below the desired accuracy ($ \bm{\varDelta}_{n+1} \leq \bm{\varDelta}_0 $) and calculates a new step size for the next step using
\begin{equation}
\delta t_{\text{new}} =
\begin{dcases}
S \, \delta t_n \, \max_i \left( \left| \frac{\varDelta_{0,i}}{\varDelta_{n+1, i}} \right| \right)^{1/5} \; \text{if } \bm{\varDelta_{n+1}} \leq \bm{\varDelta}_0 \;, \text{ or} \\
S \, \delta t_n \, \max_i \left( \left| \frac{\varDelta_{0,i}}{\varDelta_{n+1, i}} \right| \right)^{1/4} \; \text{if } \bm{\varDelta_{n+1}} > \bm{\varDelta}_0 \;.
\end{dcases}
\label{E:hnew}
\end{equation}
Here, \emph{i} represents the \emph{i}th element of the related vector and \emph{S} denotes a safety factor slightly smaller than unity (e.g., 0.9).
Equation \eqref{E:hnew} is used to calculate the next time step size both for an accepted step and also for a new, smaller step size in the case of a step rejection.
In practice, step size decreases and increases are limited to factors of ten and five, respectively~\cite{Press:1992}.

\subsection{Runge--Kutta--Chebyshev}
\label{subsec:rkc}

For ODEs with moderate levels of stiffness, one alternative to implicit algorithms is a stabilized explicit scheme such as the RKC method~\cite{Houwen:1980,Verwer:1990tg,vanderHouwen:1996ti,Verwer:1996vo,Sommeijer:1997uv,Verwer:2004gf}.
While the RKC method is explicit, it handles stiffness through additional stages---past the first two required for second-order accuracy---that extend its stability domain along the negative real axis of eigenvalues.

Our RKC implementation is taken from Sommeijer et al.~\cite{Sommeijer:1997uv} and Verwer et al.~\cite{Verwer:2004gf}.
Following the same terminology as above, the formulas for the second-order RKC are
{\allowdisplaybreaks \begin{IEEEeqnarray}{rCl}
\vec{w}_0 & = & \vec{y}_n \;, \label{E:rkc0} \\
\vec{w}_1 & = & \vec{w}_0 + \tilde{\mu}_1 \, \delta t_n \, \vec{f}_0 \;, \label{E:rkc1} \\
\vec{w}_j & = & (1 - \mu_j - \nu_j ) \vec{w}_0 + \mu_j \vec{w}_{j - 1} \nonumber \\
& & +\: \nu_j \vec{w}_{j - 2} + \tilde{\mu}_j \, \delta t_n \, \vec{f}_{j - 1} + \tilde{\gamma}_j \, \delta t_n \, \vec{f}_0, \quad j = 2, \dotsc, s \;,  \label{E:rkcj} \\
\vec{y}_{n + 1} & = & \vec{w}_s \;, \label{E:rkcs}
\end{IEEEeqnarray}}%
where \emph{s} is the total number of stages and $\vec{w}_j$ are internal vectors for the stages. The right-hand-side derivative vector function, $\vec{f}_j$, is evaluated at each stage, such that $\vec{f}_j = \vec{f} (t_n + c_j \, \delta t_n , \vec{w}_j, \vec{g} )$.
The recursive nature of $\vec{w}_j$ allows the use of only five arrays for storage.
The coefficients used in Eqs.~\eqref{E:rkc1} and \eqref{E:rkcj} can be obtained analytically for any number of stages $s \geq 2$:
\begin{align}
\tilde{\mu}_1 &= b_1 \omega_1, \quad \mu_j = \frac{2 b_j \omega_0}{b_{j - 1}} ,\quad \nu_j = \frac{-b_j}{b_{j-2}}, \quad \tilde{\mu}_j = \frac{2 b_j \omega_1}{b_{j-1}} \;, \\
\tilde{\gamma}_j &= -a_{j-1} \tilde{\mu_j} ,\quad b_0 = b_2, \quad b_1 = \frac{1}{\omega_0}, \quad b_j = \frac{T_j^{\prime \prime} (\omega_0) }{ \left( T_j^{\prime} (\omega_0) \right)^2 } \;, \\
w_0 &= 1 + \frac{\kappa}{s^2} , \quad \omega_1 = \frac{ T_s^{\prime} (\omega_0) }{ T_s^{\prime \prime} (\omega_0) } ,\quad a_j = 1 - b_j T_j \left( \omega_0 \right) \;,
\end{align}
for $j = 2, \dotsc, s$, where $\kappa \geq 0$ is the damping parameter (e.g., $\kappa = 2 / 13$~\cite{Sommeijer:1997uv,Verwer:2004gf}).
The Chebyshev polynomials of the first kind, $T_j(x)$ with first and second derivatives $T_j^{\prime}(x)$ and $T_j^{\prime \prime}(x)$, respectively, are defined recursively as
\begin{equation}
T_j (x) = 2 x T_{j - 1} (x) - T_{j - 2} (x), \quad j = 2, \dotsc, s \;,
\end{equation}
where $T_0 (x) = 1$ and $T_1 (x) = x$.
The $c_j$ used in the function evaluations are
\begin{align}
c_1 &= \frac{c_2}{T_2^{\prime}(\omega_0)} \approx \frac{c_2}{4} \;, \\
c_j &= \frac{T_s^{\prime} (\omega_0)}{T_s^{\prime \prime} (\omega_0)} \frac{T_j^{\prime \prime} (\omega_0)}{T_j^{\prime} (\omega_0)} \approx \frac{j^2 - 1}{s^2 - 1}, \quad 2 \leq j \leq s - 1 \;, \\
c_s &= 1 \;.
\end{align}

As with the RKCK method in Sec.~\ref{subsec:rkck}, the RKC method can also be used with an adaptive time stepping method for error control, as given by Sommeijer et al.~\cite{Sommeijer:1997uv}.
The error accrued in taking the step $t_{n+1} = t_n + \delta t_n$ and obtaining $\vec{y}_{n+1}$ is estimated using
\begin{equation}
\bm{\varDelta}_{n+1} = \frac{4}{5} (\vec{y}_n - \vec{y}_{n+1}) + \frac{2}{5} \delta t_n (\vec{f}_n + \vec{f}_{n+1}) \;.
\end{equation}
We then obtain the weighted RMS norm of error using this error estimate with absolute and relative tolerances:
\begin{align}
\| \bm{\varDelta}_{n+1} \|_{\text{RMS}} &= \left \| \frac{\bm{\varDelta}_{n+1}}{\vec{T} \sqrt{N}} \right \|_2 \;, \label{E:rmsE} \\
\vec{T} &= \vec{A} + R \cdot \max \left( | \vec{y}_n |, | \vec{y}_{n+1} | \right) \;,
\end{align}
where \emph{N} represents the number of unknown variables as defined previously, $\vec{A}$ is the vector of absolute tolerances, and \emph{R} is the relative tolerance.
The norm $\| \cdot \|_2$ indicates the Euclidean or $L_2$ norm.
If $ \| \bm{\varDelta}_{n+1} \|_{\text{RMS}} \leq 1 $, the step is accepted; otherwise, it is rejected and repeated using a smaller step size.
Finally, a new step size is calculated using the weighted RMS norm of error for the current and prior steps, as well as the associated step sizes, via
\begin{align}
\delta t_{n+1} &= \min \left( 10, \max( 0.1, f ) \right) \delta t_n \;, \label{eq:dtnp1} \\
f &= 0.8 \left( \frac{ \| \bm{\varDelta}_n \|_{\text{RMS}}^{1 / (p + 1)} }{ \| \bm{\varDelta}_{n+1} \|_{\text{RMS}}^{1 / (p + 1)} } \frac{\delta t_n}{\delta t_{n - 1}} \right) \frac{1}{ \| \bm{\varDelta}_n \|_{\text{RMS}}^{1 / (p + 1)} } \label{eq:rkc-f} \;,
\end{align}
where $p = 2$, the order of the algorithm.
We use Eq.~\eqref{eq:dtnp1} with a modified relation to calculate a new step size for a step rejection:
\begin{equation}
f = \frac{0.8}{ \| \bm{\varDelta}_n \|_{\text{RMS}}^{1 / (p + 1)} } \;.
\end{equation}

Determining the initial time step size requires special consideration.
First, the algorithm takes a tentative integration step, using the inverse of the spectral radius $\sigma$---the magnitude of the largest eigenvalue---of the Jacobian matrix as the step size.
Then, after estimating the error associated with this tentative step, it calculates a new step size following a similar procedure to that given in Eqs.~\eqref{eq:dtnp1} and \eqref{eq:rkc-f}:
\begin{align}
\delta t_0 &= \frac{1}{ \sigma } \;, \\
\bm{\varDelta}_0 &= \delta t_0 \left( \vec{f}(t_0 + \delta t_0, \vec{y}_0 + \delta t_0 \, \vec{f}(t_0, \vec{y}_0)) - \vec{f} (t_0, \vec{y}_0) \right) \;, \\
\delta t_1 &= 0.1 \frac{\delta t_0}{ \| \bm{\varDelta_0} \|_{\text{RMS}}^{1/2} } \;,
\end{align}
where $\|\vec{\bm{\varDelta}_0}\|_{\text{RMS}}$ is evaluated in the same manner as $\| \vec{\bm{\varDelta}}_{n+1} \|_{\text{RMS}}$ using Eq.~\eqref{E:rmsE}.

After selecting the optimal time step size to control local error, the algorithm next determines the optimal number of RKC stages in order to remain stable.
Due to stiffness, too few stages could lead to instability; in contrast, using more stages than required would add unnecessary computational effort.
The number of stages is determined using the spectral radius and time step size: 
\begin{equation}
s = 1 + \sqrt{1 + 1.54 \, \delta t_n \, \sigma} \;,
\end{equation}
as suggested by Sommeijer et al.~\cite{Sommeijer:1997uv}, where the value 1.54 is related to the stability boundary of the algorithm.
Note that \emph{s} may vary between time steps due to a changing spectral radius and time step size.
We recommend using a nonlinear power method~\cite{Sommeijer:1997uv} to calculate the spectral radius with our RKC implementation; this choice costs an additional vector to store the computed eigenvector, but avoids storing or calculating the Jacobian matrix directly.
Following Sommeijer et al.~\cite{Sommeijer:1997uv}, our RKC implementation estimates the spectral radius every 25 (internal) steps or after a step rejection.

\section{Source Code}
\label{sec:code}

Next, we provide implementation details and source code for the GPU versions of the algorithms described above.
The number of unknowns (and corresponding equations) \emph{N} is represented with the variable \lstinline+NEQN+, and the number of ODE systems $N_{\text{ode}}$ is defined as \lstinline+numODE+ in the following code.
In order for the GPU algorithms to offer a performance increase over CPU algorithms, $N_{\text{ode}}$ should be relatively large.
Although the exact number of ODEs where the GPU algorithm becomes faster than its CPU equivalent is problem dependent, Niemeyer and Sung~\cite{Niemeyer:2014dn} showed that a GPU implementation of the RKCK algorithm for chemical kinetics outperforms an equivalent serial CPU version for as little as 128 ODE systems. All operations are given here in double precision, although depending on the particular needs of the specific application single-precision calculations may be preferable to reduce the computational expense.

In order to take advantage of global memory coalescing on the GPU, we recommend storing the the set of dependent variable vectors $\vec{y}_i$, where $i = 1, \dotsc, N_{\text{ode}}$, in a single one-dimensional array, where variables corresponding to the same unknown for consecutive systems sit adjacent in memory.
In other words, if $\vec{Y}$ is a two-dimensional matrix with $N_{\text{ode}}$ rows and \emph{N} columns, where the \emph{i}th row contains the unknown vector $\vec{y}_i$, then $\vec{Y}$ should be stored in memory as a one-dimensional array in column-major ordering.
This ensures that adjacent GPU threads in the same warp access adjacent global memory locations when reading or writing equivalent array elements.
See Kirk and Hwu~\cite{Kirk:2010we} or Jang et al.~\cite{Jang:2011ct} for more details and examples on global memory coalescing.

The following code snippet shows the proper loading of a host array \lstinline+yHost+ from an arbitrary array \lstinline+y+ that contains initial conditions for all ODEs:
\begin{lstlisting}
double *yHost;
yHost = (double *) malloc (numODE * NEQN * sizeof(double));

for (int i = 0; i < numODE; ++i) {
    for (int j = 0; j < NEQN; ++j) {
      yHost[i + numODE * j] = y[i][j];
	}
}
\end{lstlisting}
A similar procedure should be used for the constant parameter vector $\vec{g}$ if needed.

Next, the GPU global memory arrays should be declared and initialized, and the block\slash grid dimensions set up using
\begin{lstlisting}
double *yDevice;
cudaMalloc ((void**) &yDevice, numODE * NEQN * sizeof(double));

int blockSize;
if (numODE < 4194304) {
	blockSize = 64;
} else if (numODE < 8388608) {
	blockSize = 128;
} else if (numODE < 16777216) {
	blockSize = 256;
} else {
	blockSize = 512;
}
dim3 dimBlock (blockSize, 1);
dim3 dimGrid (numODE / dimBlock.x, 1);
\end{lstlisting}
Here, we use simple one-dimensional block and grid dimensions; reshaping the grid should not affect performance, but can be done for convenience.
We chose 64 as the block size for problems with less than \num{4194304} ODEs based on experience for chemical kinetics problems~\cite{Niemeyer:2014dn}.
The size should remain a power of two, but different block sizes may be optimal for other problems.

The final step is to set up the ODE integration loop and kernel function execution.
The integration driver kernel, to be described shortly, will perform internal sub-stepping as necessary to reach the specified end time.
Depending on the objectives, there are various ways to approach this:
\begin{itemize}
\item If only the final integrated results are needed, then a single GPU kernel can be invoked.
\item If intermediate integration results are needed, then an acceptable outer step size over which results will be spaced should be chosen, and the GPU kernel should be invoked inside a loop.
\end{itemize}
We will leave the code as general as possible by following the second approach, although modifications should be made depending on the desired functionality.
In both cases, the global memory array holding the variables to be integrated needs to be transferred to the GPU before, and from the GPU after, each kernel invocation.
\begin{lstlisting}
// set initial time
double t = t0;
double tNext = t + h;

while (t < tEnd) {
	// transfer memory to GPU
	cudaMemcpy (yDevice, yHost, numODE*NEQN*sizeof(double), cudaMemcpyHostToDevice);
	
	intDriver <<<dimGrid, dimBlock>>> (t, tNext, numODE, gDevice, yDevice);
	
	// transfer memory back to CPU
	cudaMemcpy (yHost, yDevice, numODE*NEQN*sizeof(double), cudaMemcpyDeviceToHost);
	
	t = tNext;
	tNext += h;
}

cudaFree (gDevice);
cudaFree (yDevice);
\end{lstlisting}
Here, \lstinline+t0+ refers to the initial time, \lstinline+tEnd+ the desired final time, and \lstinline+h+ the outer step size.
In the current form, each outer integration step performed by the GPU will be a ``restart'' integration, meaning no information about previous steps (e.g., error estimates, step sizes) will be used to assist the startup.
This is necessary in certain applications such as reactive-flow simulations (and other simulation methods that use operator splitting), where after each outer step integration results are combined with changes due to fluid transport, thereby invalidating stored integration information.
However, where possible, better performance may be obtained by transferring the appropriate data from the GPU and used in the next overall integration step.

The next code snippet contains the general integration driver kernel, suitable for either algorithm:
\begin{lstlisting}
__global__ void
intDriver (const double t, const double tEnd, const int numODE,
           const double* gGlobal, double* yGlobal) {
	// unique thread ID, based on local ID in block and block ID
	int tid = threadIdx.x + (blockDim.x * blockIdx.x);
	
	// ensure thread within limit
	if (tid < numODE) {
	
	// local array with initial values
	double yLocal[NEQN];
	
	// constant parameter(s)
	double gLocal = gGlobal[tid];
	
	// load local array with initial values from global array
	for (int i = 0; i < NEQN; ++i) {
		yLocal[i] = yGlobal[tid + numODE * i];
	}
	
	// call integrator for one time step
	integrator (t, tEnd, yLocal, gGlobal);
	
	// update global array with integrated values
	for (int i = 0; i < NEQN; ++i) {
		yGlobal[tid + numODE * i] = yLocal[i];
	}	
	}
}
\end{lstlisting}
The function \lstinline+integrator+ should be replaced with \lstinline+rkckDriver+ or \lstinline+rkcDriver+ (given below) depending on the desired integration algorithm.

\subsection{RKCK Code}
\label{sec:code-rkck}

In the following, the source code for the RKCK driver device function is given in functional snippets.
First, the minimum and maximum allowable time step sizes are defined, and the initial step size is set as half the integration time.
\begin{lstlisting}
__device__ void
rkckDriver (double t, const double tEnd, const double g,
            double* y) {
	
	// maximum and minimum allowable step sizes
	const double hMax = fabs(tEnd - t);
	const double hMin = 1.0e-20;
	
	// initial step size
	double h = 0.5 * fabs(tEnd - t);
\end{lstlisting}

Then, inside the time integration loop, the algorithm takes a trial integration step and estimates the error of that step.
\begin{lstlisting}
	// integrate until specified end time
	while (t < tEnd) {
	
		// limit step size based on remaining time
		h = fmin(tEnd - t, h);
		
		// y and error vectors temporary until error determined	
		double yTemp[NEQN], yErr[NEQN];
		
		// evaluate derivative
		double F[NEQN];
		dydt (t, y, g, F);
		
		// take a trial step
		rkckStep (t, y, g, F, h, yTemp, yErr);
		
		// calculate error
		double err = 0.0;
		int nanFlag = 0;
		for (int i = 0; i < NEQN; ++i) {
			if (isnan(yErr[i])) nanFlag = 1;

			err = fmax(err, fabs(yErr[i] / (fabs(y[i]) + fabs(h * F[i]) + TINY)));
		}
		err /= eps;
\end{lstlisting}

If the error is too large, the step size is decreased and the step retaken; otherwise, the algorithm accepts the step and calculates the next step size, then repeats the process.
\begin{lstlisting}
		// check if error too large
		if ((err > 1.0) || isnan(err) || (nanFlag == 1)) {
			// step failed, error too large
			if (isnan(err) || (nanFlag == 1)) {
				h *= P1;
			} else {
				h = fmax(SAFETY * h * pow(err, PSHRNK), P1 * h);
			}
			
		} else {
			// step accepted
			t += h;
		
			if (err > ERRCON) {
				h = SAFETY * h * pow(err, PGROW);
			} else {
				h *= 5.0;
			}
			
			// ensure step size is bounded
			h = fmax(hMin, fmin(hMax, h));
		
			for (int i = 0; i < NEQN; ++i)
				y[i] = yTemp[i];
		}
	}
}
\end{lstlisting}
The device function \lstinline+dydt+ evaluates the derivative function \lstinline+F+ for the particular problem as in Eq.~\eqref{eq:ode} using the input time \lstinline+t+, vector of dependent variables \lstinline+y+, and constant parameter(s) \lstinline+g+.
The device function \lstinline+rkcStep+, not given here, takes a single integration step using Eqs.~\eqref{eq:k1}--\eqref{eq:rkc-ynp1}, returning the vector of integrated values \lstinline+yTemp+ as well as the error vector \lstinline+yErr+.
A number of constants were used in this function, given here:
\begin{lstlisting}
#define UROUND (2.22e-16)
#define SAFETY 0.9
#define PGROW (-0.2)
#define PSHRNK (-0.25)
#define ERRCON (1.89e-4)
#define TINY (1.0e-30)
const double eps = 1.0e-10;
\end{lstlisting}

\subsection{RKC Code}
\label{sec:code-rkc}

The RKC driver algorithm is next given.
For this algorithm, the number of stages must be determined at each step to handle local stiffness; to avoid excess computation, a maximum number of stages is first set.
In addition, minimum and maximum allowable time step sizes are defined.
\begin{lstlisting}
__device__ void
rkcDriver (double t, const double tEnd, const double g, double* y) {
	// number of steps
	int numStep = 0;
	
	// maximum allowable number of RKC stages
	int mMax = (int)(round(sqrt(relTol / (10.0 * UROUND))));
	
	// RKC needs at least two stages for second-order accuracy
	if (mMax < 2) mMax = 2;
	
	// maximum allowable step size
	const double stepSizeMax = fabs(tEnd - t);
	
	// minimum allowable step size
	double stepSizeMin = 10.0*UROUND*fmax(fabs(t), stepSizeMax);
\end{lstlisting}

Then, the algorithm evaluates the derivative using the initial conditions for use as the initial eigenvector estimate for the spectral radius calculation.
The calculated eigenvectors are stored and used as initial guesses in later steps.
\begin{lstlisting}
	// internal y vector
	double y_n[NEQN];
	for (int i = 0; i < NEQN; ++i)
		y_n[i] = y[i];
	
	// calculate F_n for initial y
	double F_n[NEQN];
	dydt (t, y_n, g, F_n);
	
	// internal work vector
	double work[4 + NEQN];
	
	// load initial estimate for eigenvector
	if (work[2] < UROUND) {
		for (int i = 0; i < NEQN; ++i) {
			work[4 + i] = F_n[i];
		}
	}
\end{lstlisting}

Inside the time integration loop, the algorithm calculates the spectral radius for the first step---which it next uses to determine the initial step size---and every 25 steps thereafter.
\begin{lstlisting}
	// perform internal sub-stepping
	while (t < tEnd) {
		double tempArr[NEQN], tempArr2[NEQN], err;
		
		// if last step, limit step size
		if ((1.1 * work[2]) >= fabs(tEnd - t)) work[2] = fabs(tEnd - t);

		// estimate Jacobian spectral radius if 25 steps passed
		if ((numStep % 25) == 0) {
			work[3] = rkcSpecRad (t, y_n, g, F_n, stepSizeMax, &work[4], tempArr2);
		}
\end{lstlisting}

For the initial step, a trial step is taken using the inverse of the spectral radius as the step size; the resulting error is used to determine an appropriate step size that satisfies error control.
\begin{lstlisting}
		// if this is initial step
		if (work[2] < UROUND) {
			// estimate first time step
			work[2] = stepSizeMax;
			
			if ((work[3] * work[2]) > 1.0) work[2] = 1.0 / work[3];
			work[2] = fmax(work[2], stepSizeMin);

			for (int i = 0; i < NEQN; ++i) {
				temp_arr[i] = y_n[i] + (work[2] * F_n[i]);
			}
			dydt (t + work[2], tempArr, g, tempArr2);
			
			err = 0.0;
			for (int i = 0; i < NEQN; ++i) {
				double est = (tempArr2[i] - F_n[i]) / (absTol + relTol * fabs(y_n[i]));
				err += est * est;
			}
			err = work[2] * sqrt(err / NEQN);

			if ((P1 * work[2]) < (stepSizeMax * sqrt(err))) {
				work[2] = fmax(P1 * work[2] / sqrt(err), stepSizeMin);
			} else {
				work[2] = stepSizeMax;
			}
		}
\end{lstlisting}
For all steps following the first, the value stored in \lstinline+work[2]+ is used for the time step size.

Next, the number of stages is determined using the spectral radius and current time step size, and a tentative integration step performed.
\begin{lstlisting}
		// calculate number of steps
		int m = 1 + (int)(sqrt(1.54 * work[2] * work[3] + 1.0));
		
		// modify step size based on stages
		if (m > mMax) {
			m = mMax;
			work[2] = ((double)(m * m - 1)) / (1.54 * work[3]);
		}
		
		// perform tentative time step
		rkcStep (t, y_n, g, F_n, work[2], m, y);
\end{lstlisting}

The algorithm then estimates the error of that step.
\begin{lstlisting}
		// calculate derivative F_np1 with tentative y_np1
		dydt (t + work[2], y, g, tempArr);
		
		// estimate error
		err = 0.0;
		for (int i = 0; i < NEQN; ++i) {
			double est = 0.8 * (y_n[i] - y[i]) + 0.4 * work[2] * (F_n[i] + tempArr[i]);
			est /= (absTol + relTol * fmax(fabs(y[i]), fabs(y_n[i])));
			err += est * est;
		}
		err = sqrt(err / ((double)N));
\end{lstlisting}

Based on the error magnitude, the algorithm determines whether to accept the step and proceed to the next step or decrease the step size and repeat the current step.
\begin{lstlisting}
		// check value of error
		if (err > 1.0) {
			// error too large, step is rejected
			// select smaller step size
			work[2] = 0.8 * work[2] / (pow(err, (1.0 / 3.0)));
			
			// reevaluate spectral radius
			work[3] = rkcSpecRad (t, y_n, g, F_n, stepSizeMax, &work[4], tempArr2);
		} else {
			// step accepted
			t += work[2];
			numStep++;
\end{lstlisting}			

Finally, for an accepted step, the current step size and error are stored and the next step size is calculated.
\begin{lstlisting}
			double fac = 10.0;
			double temp1, temp2;
			
			if (work[1] < UROUND) {
				temp2 = pow(err, (1.0 / 3.0));
				if (0.8 < (fac * temp2)) fac = 0.8 / temp2;
			} else {
				temp1 = 0.8 * work[2] * pow(work[0], (1.0 / 3.0));
				temp2 = work[1] * pow(err, (2.0 / 3.0));
				if (temp1 < (fac * temp2)) fac = temp1 / temp2;
			}

			// set "old" values to those for current time step
			work[0] = err;
			work[1] = work[2];

			for (int i = 0; i < NEQN; ++i) {
				y_n[i] = y[i];
				F_n[i] = tempArr[i];
			}
			
			work[2] *= fmax(P1, fac);
			work[2] = fmax(stepSizeMin, fmin(stepSizeMax, work[2]));
		}
	}
}
\end{lstlisting}
As before, we do not provide the RKC integration step device function \lstinline+rkcStep+, which evaluates Eqs.~\eqref{E:rkc0}--\eqref{E:rkcs}.
The absolute and relative tolerances \lstinline+absTol+ and \lstinline+relTol+ are set as defined constants, e.g.,:
\begin{lstlisting}
const double abs_tol = 1.0e-10;
const double rel_tol = 1.0e-6;
\end{lstlisting}
Note that these may be modified to more stringent tolerances if desired.
The constant \lstinline+UROUND+ is defined the same as in the RKCK code above.
The local work array \lstinline+work+ contains, in element order:
\begin{enumerate}
\item[0] the previous step error estimation,
\item[1] previous time step,
\item[2] current time step,
\item[3] spectral radius, and
\item[4] vector of eigenvalues (of size \emph{N}).
\end{enumerate}
The device function \lstinline+rkcSpecRad+ returns the spectral radius, the largest magnitude eigenvalue; various methods may be used for this purpose depending on the case.
We provide GPU source code for a nonlinear power method adopted from Sommeijer et al.~\cite{Sommeijer:1997uv} that may be used for general applications in the \nameref{sec:appendix}.

\section{Performance Results}
\label{sec:results}

We tested the performance of the GPU-based RKCK and RKC integration algorithms using two ODE test cases, ranging the number of ODE systems from \numrange{e1}{e5}.
For both cases, all calculations were performed in double precision using a single GPU and single CPU; we compared the performance of the GPU algorithm against serial CPU calculations as well as parallelized CPU performance---via OpenMP~\cite{OpenMP:2008}---on four cores.
We performed the GPU calculations using an NVIDIA Tesla c2075 GPU with \SI{6}{\giga\byte} of global memory, and an Intel Xeon X5650 CPU, running at \SI{2.67}{\giga\hertz} with \SI{256}{\kilo\byte} of L2 cache memory per core and \SI{12}{\mega\byte} of L3 cache memory, served as the host processor both for the GPU calculations and the CPU single- and four-core OpenMP calculations.
We used the GNU Compiler Collection (gcc) version 4.6.2 (with the compiler options ``\texttt{-O3 -ffast-math -std=c99 -m64}'') to compile the CPU programs and the CUDA 5.5 compiler nvcc version 5.5.0 (``\texttt{-O3 -arch=sm\_20 -m64}'') to compile the GPU versions.
We set the GPU to persistence mode, but also used the \lstinline+cudaSetDevice()+ to hide any further device initialization delay in the CUDA implementations prior to the timing.

The integration algorithms take as input initial conditions and a global time step, performing internal sub-stepping as necessary.
The computational wall-clock times reported represent the average over ten global time steps, which for the GPU versions includes the overhead required for transmitting data between the CPU and GPU before and after each global step.
The integrator restarts at each global time step, not storing any data from the previous step---although any sub-stepping performed by the algorithm within these larger steps does benefit from retained information from prior sub-steps.
Interested readers should refer to Niemeyer and Sung~\cite{Niemeyer:2014dn} for more detailed performance evaluations of these algorithms in the context of chemical kinetics problems.

\subsection{RKCK Results}
\label{subsec:results-rkck}

We used the Pleiades ODE test problem (PLEI) of Hairer et al.~\cite{Hairer:1993,Mazzia:2008} to test the GPU- and CPU-based versions of the RKCK integrator.
This nonstiff test case originates from a celestial mechanisms problem tracking the coordinates of seven stars; it consists of a set of 14 second-order ODEs based on Newtonian gravitational forces, in the form
{\allowdisplaybreaks \begin{IEEEeqnarray}{rCl}
\vec{z}^{\prime \prime} &=& \binom{\vec{x}}{\vec{y}}^{\prime\prime} = \binom{\vec{f}^{(1)} \left( \vec{x}, \vec{y} \right)}{\vec{f}^{(2)} \left( \vec{x}, \vec{y} \right)} , \quad \vec{z} \in \Re^{14} \;, \\
x_{i}^{\prime\prime} &=& f_{i}^{(1)} \left( \vec{x}, \vec{y} \right) = \sum_{j \neq i} m_j \left( x_j - x_i \right) / r_{ij} \;, \\
y_{i}^{\prime\prime} &=& f_{i}^{(2)} \left( \vec{x}, \vec{y} \right) = \sum_{j \neq i} m_j \left( y_j - y_i \right) / r_{ij} \;, \\
r_{ij} &=& \left( \left(x_i - x_j\right)^2 + \left(y_i - y_j\right) \right)^{3/2}, \quad i,j = 1,\dotsc,7 \;,
\end{IEEEeqnarray}}%
where ($x_i$, $y_i$) and $m_i = i$ are the coordinates and mass of the \emph{i}th star, respectively.
This second-order system can be converted into a system of 28 first-order ODEs of the form by defining $\vec{w} = \vec{z}^{\prime}$, such that
\begin{equation}
\binom{\vec{z}}{\vec{w}}^{\prime} = \binom{\vec{w}}{\vec{f}(\vec{z})} ,\quad
\binom{\vec{z}}{\vec{w}} \in \Re^{28} \;.
\end{equation}

While the original problem offers specific initial conditions for a single ODE system, here we consider a large number of ODEs with the initial conditions randomly perturbed by a small factor to emulate a sensitivity analysis.
We integrated the ODE systems from \emph{t} = \SIrange{0}{1.0}{\second} using \SI{1.0e-1}{\second} as the global time step size.
We set the RKCK tolerance $\varepsilon$ (\lstinline+eps+ in the code) to \num{1.0e-10}.

\begin{figure}[tbp]
\begin{center}
\includegraphics[width=0.8\linewidth]{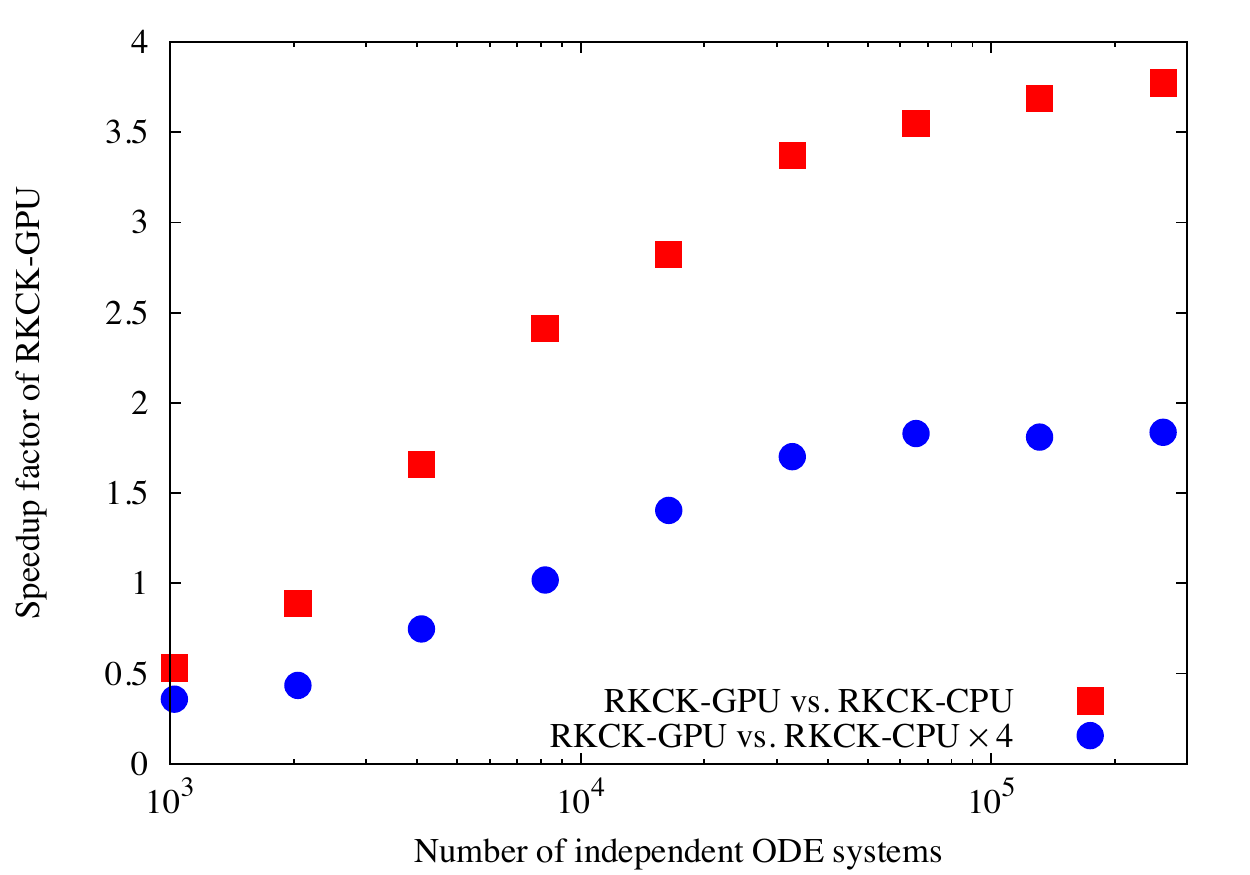}
\caption{Speedup factors offered by GPU-based explicit RKCK integration algorithm over single- and four-core CPU-based versions for Pleiades ODE problem. Note that the horizontal axis is displayed in logarithmic scale.}
\label{fig:rkck}
\end{center}
\end{figure}

Figure~\ref{fig:rkck} shows the speedup factors, measured as the ratio of computational times per step, offered by the GPU-based RKCK algorithm over the baseline CPU version for both serial and four-core parallel operation, for numbers of ODE systems ranging from \numrange{1024}{262144}.
The GPU-based algorithm ran faster than the serial and parallel CPU-based algorithms for $N_{\text{ode}}$ larger than \num{4096} and \num{8192}, respectively.
For the current problem, at best the GPU offered a speedup factors of nearly four and two over the serial and four-core CPU implementations, respectively.
The non-smooth performance scaling resulted from the randomly perturbed initial conditions.

Note that since each ODE system used randomly perturbed initial conditions, adjacent threads in the GPU implementation handled potentially extremely different initial condition values, resulting in thread divergence due to varying internal time step sizes.
Therefore, the results shown here represent a worst-case GPU algorithm performance, particularly compared to applications involving operator-split reactive-flow codes where adjacent threads\slash ODE systems correspond to neighboring spatial locations.
In such situations, initial conditions would be more similar and therefore follow similar instruction pathways.
In either case, GPU-based integration algorithms offer performance benefits over the baseline CPU versions.
See Niemeyer and Sung~\cite{Niemeyer:2014dn} for more discussion on this topic.

Furthermore, the current problem involved a relatively simple system of ODEs, limiting the calculations performed on the GPU between the memory transfers before and after each integration step.
ODE systems with more complex derivative functions would saturate the GPU with operations, increasing performance.
For example, the RKCK algorithm demonstrated by Niemeyer and Sung~\cite{Niemeyer:2014dn} performed up to 126 times faster on a GPU than on a serial CPU, integrating a chemical kinetics ODE system with nine species participating in 38 reaction steps---requiring significantly more floating-point calculations than the case studied here.

\subsection{RKC Results}
\label{subsec:results-rkc}

To demonstrate the performance of the GPU-based RKC algorithm, we used a chemical kinetics problem: the ODE system describing the constant-volume autoignition of ethanol (\ce{C2H5OH}).
We implemented the reaction mechanism of Marinov~\cite{Marinov:1999} to describe the oxidation of ethanol, with 57 species participating in 766 irreversible reaction steps.
The governing ODE system contained 58 equations: one for temperature \emph{T} and the rest for species mass fractions $\vec{Y}$:
{\allowdisplaybreaks \begin{IEEEeqnarray}{rCl}
\frac{d \vec{y}}{d t} &=& \left( \frac{d T}{d t}, \frac{dY_1}{dt}, \dotsc, \frac{d Y_{N_{\text{sp}}}}{dt} \right)^{\intercal} \;, \\
\frac{dT}{dt} &=& -\frac{1}{\rho c_v} \sum_{i = 1}^{N_{\text{sp}}} e_i \omega_i W_i \;, \\
\frac{dY_i}{dt} &=& \frac{W_i \omega_i}{\rho} , \quad i = 1, \dotsc, N_{\text{sp}} \;, \\
\omega_i &=& \sum_{j=1}^{N_{\text{reac}}} \left( \nu_{ij}^{\prime\prime} - \nu_{ij}^{\prime} \right) k_j \prod_{k=1}^{N_{\text{sp}}} C_k^{\nu_{kj}^{\prime}} \;,
\end{IEEEeqnarray}}%
where $\rho$ indicates the density, $c_v$ the mass-averaged constant-volume specific heat, $e_i$ the internal energy of the \emph{i}th species, $W_i$ the molecular weight of the \emph{i}th species, $\nu_{ij}^{\prime\prime}$ and $\nu_{ij}^{\prime}$ the forward and reverse stoichiometric coefficients for the \emph{i}th species in reaction \emph{j}, $C_k$ the molar concentration of the \emph{k}th species, and $N_{\text{sp}}$ and $N_{\text{reac}}$ are the numbers of species and reactions, respectively.
For a reaction \emph{j} without pressure dependence, the rate coefficient $k_j$ is given in Arrhenius form by
\begin{equation}
k_j = A_j T^{\beta_j} \exp \left( \frac{-E_j}{\mathcal{R} T} \right) \;, 
\end{equation}
where $\mathcal{R}$ is the universal gas constant, $A_j$ the pre-exponential coefficient, $\beta_j$ the temperature exponent, and $E_j$ the activation energy.
Note that reactions can be pressure-dependent (see, e.g., Law~\cite{Law:2006} for examples of various pressure-dependence formulations); these were also considered in the current implementation.

This problem is moderately stiff using a time step size of $\delta t$ = \SI{1.0e-6}{\second} for 10 global time steps.
In this case, we generated initial conditions for the set of ODE systems by sampling the solutions obtained from constant-pressure homogeneous ignition simulations, initiated at \SI{1600}{\kelvin}, \SI{1}{\atm}, and an equivalence ratio of one\footnote{An equivalence ratio of one indicates the mixture of fuel and oxidizer set to an appropriate ratio for complete combustion.}.
We assigned these initial conditions sequentially, such that adjacent threads in the GPU implementation contained data from consecutive time steps in the sample---and therefore such threads handled the integration of similar conditions, emulating adjacent spatial locations in an operator-split reactive-flow simulation.

\begin{figure}[tbp]
\begin{center}
\includegraphics[width=0.8\linewidth]{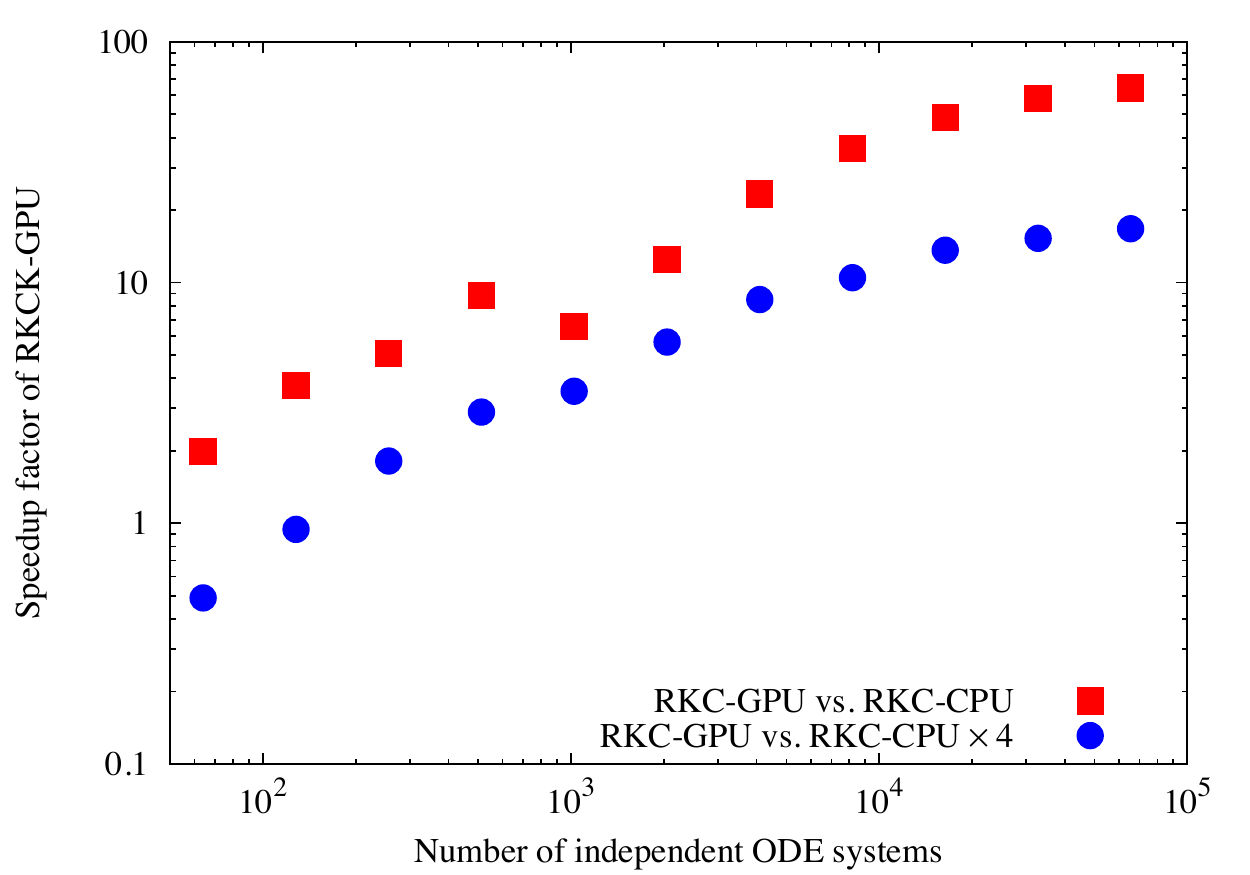}
\caption{Speedup factors offered by GPU-based explicit RKC integration algorithm over single- and four-core CPU-based versions for pollution ODE problem. Note that both axes are displayed in logarithmic scale.}
\label{fig:rkc}
\end{center}
\end{figure}

Figure~\ref{fig:rkc} shows the speedup factors offered by the GPU-based RKC algorithm over the baseline CPU version for  both serial and four-core parallel operation, for numbers of ODE systems ranging from \numrange{64}{16384}.
In this case, the GPU-accelerated code ran faster than the serial CPU version for the entire range of ODE system sizes considered, while it offered better performance than the four-core parallel CPU version for 256 ODEs and higher.
At best, the GPU-based RKC algorithm ran nearly 64 and 17 times faster than the serial and four-core CPU implementations, respectively.
The discontinuity in speedup seen in Fig.~\ref{fig:rkc} corresponded to the inclusion of conditions leading to greater stiffness.

\section{Conclusions}
\label{sec:conclusions}

In this chapter, we presented two explicit algorithms appropriate for integrating large numbers of independent ODE systems on GPUs.
Specifically, we proposed the fifth-order adaptive Runge--Kutta--Cash--Karp (RKCK) method for nonstiff problems and the stabilized second-order adaptive Runge--Kutta--Chebyshev (RKC) method for problems with moderate levels of stiffness.
Source code and implementation details were presented to ease the adoption of such methods, and performance comparison results were presented for each method.
The examples shown here served to demonstrate the potential of GPU acceleration where many independent systems of ODEs need to be integrated; in the case of the RKC algorithm, we demonstrated more than an order of magnitude performance increase over an equivalent parallel CPU code running on four cores.
The types of scientific and engineering problems dealing with large numbers of ODEs---in particular, reactive-flow models that rely on operator splitting---can benefit significantly from GPU acceleration; interested readers can directly implement the algorithms presented here to such ends, or use them as the beginnings for their own solution.

\section*{Acknowledgements}
This work was supported by the US Department of Defense through the National Defense Science and Engineering Graduate Fellowship program, the National Science Foundation Graduate Research Fellowship under grant number DGE-0951783, and the Combustion Energy Frontier Research Center---an Energy Frontier Research Center funded by the US Department of Energy, Office of Science, Office of Basic Energy Sciences under award number DE-SC0001198.

\section*{Appendix}
\label{sec:appendix}
\addcontentsline{toc}{section}{Appendix}

Various methods may be used to calculate the spectral radius, including the Gershgorin circle theorem~\cite{Gersgorin:1931,Horn:1990} that provides an upper-bound estimate.
Here, we provide a function based on a nonlinear power method~\cite{Sommeijer:1997uv}.
\begin{lstlisting}
__device__ double
rkcSpecRad (const double t, const double* y, const double g, const double* F, const double hMax, double* v, double* Fv) {
	// maximum number of iterations
	const int itmax = 50;
	
	double small = 1.0 / hmax;
	
	double nrm1 = 0.0;
	double nrm2 = 0.0;
	for (int i = 0; i < NEQN; ++i) {
		nrm1 += (y[i] * y[i]);
		nrm2 += (v[i] * v[i]);
	}
	nrm1 = sqrt(nrm1);
	nrm2 = sqrt(nrm2);

	double dynrm;
	if ((nrm1 != 0.0) && (nrm2 != 0.0)) {
		dynrm = nrm1 * sqrt(UROUND);
		for (int i = 0; i < NEQN; ++i) {
			v[i] = y[i] + v[i] * (dynrm / nrm2);
		}
	} else if (nrm1 != 0.0) {
		dynrm = nrm1 * sqrt(UROUND);
		for (int i = 0; i < NEQN; ++i) {
			v[i] = y[i] * (1.0 + sqrt(UROUND));
		}
	} else if (nrm2 != 0.0) {
		dynrm = UROUND;
		for (int i = 0; i < NEQN; ++i) {
			v[i] *= (dynrm / nrm2);
		}
	} else {
		dynrm = UROUND;
		for (int i = 0; i < NEQN; ++i) {
			v[i] = UROUND;
		}
	}

	// now iterate using nonlinear power method
	double sigma = 0.0;
	for (int iter = 1; iter <= itmax; ++iter) {
		
		dydt (t, pr, v, Fv);

		nrm1 = 0.0;
		for (int i = 0; i < NEQN; ++i) {
			nrm1 += ((Fv[i] - F[i]) * (Fv[i] - F[i]));
		}
		nrm1 = sqrt(nrm1);
		nrm2 = sigma;
		sigma = nrm1 / dynrm;

		nrm2 = fabs(sigma - nrm2) / sigma;
		if ((iter >= 2) && (fabs(sigma - nrm2) <= (fmax(sigma, small) * 0.01))) {
			for (int i = 0; i < NEQN; ++i) {
				v[i] -= y[i];
			}
			return (1.2 * sigma);
		}

		if (nrm1 != 0.0) {
			for (int i = 0; i < NEQN; ++i) {
				v[i] = y[i] + ((Fv[i] - F[i]) * (dynrm / nrm1));
			}
		} else {
			int ind = (iter % NEQN);
			v[ind] = y[ind] - (v[ind] - y[ind]);
		}
	}
	return (1.2 * sigma);
}
\end{lstlisting}

\bibliographystyle{h-physrev}
\bibliography{refs}

\end{document}